\documentstyle[11pt,paspconf]{article}

\begin{document}

\title{Degeneracies of the radial mass profile in lens models}

\author{Olaf Wucknitz\altaffilmark{1} and Sjur Refsdal\altaffilmark{2}}
\affil{Hamburger Sternwarte, Gojenbergsweg 112, 21029 Hamburg,
  Germany}
\altaffiltext{1}{email: {\tt owucknitz@hs.uni-hamburg.de}}
\altaffiltext{2}{email: {\tt srefsdal@hs.uni-hamburg.de}}

\begin{abstract}
  We discuss a simple parameter degeneracy in lens models which
  follows directly from the well known mass sheet degeneracy and
  prevents determining the radial mass distribution from lensing alone
  for many systems.  For cosmological important systems like 1115+080
  it is shown that no exact value of $H_0$ can be determined without
  assuming a mass index $\beta$.
\end{abstract}


\section{The mass sheet degeneracy}

The degeneracies examined in this poster are a consequence of the so
called ``mass sheet degeneracy''. Falco et al. (1985) and Gorenstein
et al. (1988) have shown that a lens model can be transformed to an
equivalent model by scaling it with a factor of $(1-\kappa)$ and
adding a constant surface mass density $\kappa$.  Constant mass
densities as well as external shear do not contribute to the time
delay. As $H_0\,\Delta t$ depends linearly on the mass density, $H_0$
also scales with $(1-\kappa)$.  This has the effect that only an upper
limit for $H_0$ can be determined when additional mass sheets may be
present.

When modelling lens systems, no decision between different models can
be made if the models are equivalent {\em just for the positions of
  the observed images.} We want to examine the impact of the
degeneracy for simple parametric lens models.

\section{Perturbed spherical lens models}

We use the very simple approach of a spherically symmetric mass
distribution plus external shear $\gamma$ to illustrate the
degeneracy.  For a a radial deflection angle of $\alpha(r)$ we can
find an equivalent model by scaling $\alpha(r)$ and the shear by
$(1-\kappa)$ and adding the surface mass density which gives a
contribution of $\kappa\,r$ to the radial deflection angle.

Whether this leads to a degeneracy of model parameters, depends on the
number of images, the image configuration and the number of parameters
used to describe the radial deflection angle $\alpha$.

A very simple while important case is the spherical power law model
with a deflection angle of $\alpha=\alpha_0\,r^{\beta-1}$. It has two
parameters, the scale $\alpha_0$ and the mass index $\beta$. Special
values of $\beta$ are 0 (point mass), 1 (isothermal sphere) and 2
(constant mass sheet). This simple model may not describe the mass
distribution of real galaxies exactly but is quite useful to
illustrate the parameter degeneracies which exist for other models as
well.

\section{Application to Einstein cross systems}

In systems like the Einstein cross 2237+0305, the images are located
at more or less the same distance (about one Einstein radius) from the
centre of the lens. Two spherical models are equivalent to first order
if the deflection angle as well as the first derivative is the same
for both. We use the Einstein radius as unit of $r$ and an isothermal
model as reference.  A model with $\alpha = \alpha_0\,r^{\beta-1}$ is
equivalent to the isothermal model to first order near $r=1$ if
$\alpha_0=1$ and $\beta=1+\kappa$.  This leads to a very simple
scaling of the shear, the time delay and $H_0$:
\begin{equation}
1-\kappa \; = \; 2-\beta \; = \;
 \frac{\gamma_{(\beta)}}{\gamma_{\rm(iso)}} \;  = \;
 \frac{H_{0\,(\beta)}}{H_{0\,\rm(iso)}}
\end{equation}
We see that the Hubble constant for the more general $\beta$ model $H_{0\,(\beta)}$ can
differ significantly from the value determined for the isothermal
model $H_{0\,\rm(iso)}$. Even more important is the fact that the possible
systematic error which is made by assuming isothermal models when the real
$\beta$ is different is the same {\em for all systems of this type}
and does not show as scatter in the results for $H_0$.

Comparison of our simple analytical results with numerical models from
Wambsganss~\&~Paczy\'nski (1994) for the Einstein cross and Schechter
et al. (1997) for the ``triple quasar'' 1115+080 show that the
agreement is excellent for the former and still quite good for the
latter.  The case of 1115+080 is especially interesting, because it is
one of the few systems with a measured time delay used for the
determination of $H_0$ (Schechter et al. 1997).

\section{Further information}

The original much more detailed poster presented at the conference is
available from {\tt
  http://www.hs.uni-hamburg.de/english/persons/wucknitz.html} or on
request.  An article with a discussion of the degeneracy for different
types of lens systems is in preparation.

\end{document}